\documentclass[%
 reprint,
superscriptaddress,
 amsmath,amssymb,
 aps,
floatfix,
preprint, onecolumn,
]{revtex4-2}

\usepackage{graphicx}% Include figure files
\usepackage{dcolumn}% Align table columns on decimal point
\usepackage{bm}% bold math
\def\r{\bm{r}}
\def\n{\bm{\hat n}}
\def\f{\bm{f}}
\def\F{\bm{F}}
\def\rr{\bm{\zeta}}
\def\R{{\bm{R}}}
\def\d{{\mathrm{d}}}
\def\H{{\bm{H}}}
\def\G{{\bm{G}}}
\def\q{{\bm{q}}}
\def\e{{\bm{e}}}
\def\v{{\bm{v}}}
\def\T{{\bm{T}}}
\def\g{{\bm{g}}}

\newcommand{\dbar}{{\mkern5mu\mathchar'26\mkern-12mu \d}}

\begin{document}

%\preprint{APS/123-QED}

\title{Random packing in three dimensions}% Force line breaks with \\

\author{Chaoming Song}
\email{c.song@miami.edu}
\affiliation{%
Department of Physics, University of Miami, Coral Gables, Florida 33142, USA}%



%\date{}% It is always \today, today,

\begin{abstract}
Unraveling the complexities of random packing in three dimensions has long puzzled physicists. While both experiments and simulations consistently show a maximum density of 64 percent for tightly packed random spheres, we still lack an unambiguous and universally accepted definition of random packing. This paper introduces an innovative standpoint, depicting random packing as spheres closest to a set of  quenched random points. We furnish an efficacious algorithm to probe this proposed model numerically. We unearth a unique out-of-equilibrium thermodynamic phenomenon, akin to a `latent heat', that emerges at $\phi_J \approx 0.645$ in three dimensions. This phenomenon is accompanied by global and local structural rearrangements, marking a jamming transition from an unjammed state to a jammed one. Notably, such a `jamming' transition is absent for two-dimensional random packing. Our innovative approach paves a new avenue for defining random packing and provides novel insights into the behavior of amorphous materials.
\end{abstract}

\maketitle
\newpage

%\tableofcontents

\section*{Introduction}

The maximal spherical packing has a well-known packing fraction of $\phi_m = \pi/\sqrt{18} \approx 0.74$,  originally conjectured by Kepler and proven recently \cite{hales2005proof}. In contrast, the nature of Random Close Packing (RCP), a term coined by Bernal et al. in the 1960s \cite{bernal1960packing}, remains elusive. RCP refers to the densest random arrangement of spheres, with an empirically observed packing fraction of approximately $\phi_\text{RCP} \approx 0.64$. Despite over half a century's efforts to understand random packing, little progress has been made toward an unambiguous and widely-acceptable definition of RCP. The main challenge lies in the very nature of ``randomness." While it is easy to define a random set of uncorrelated points, such as a Poisson point process, it is difficult to generalize these definitions to random packing. This is because overlaps are disallowed in sphere packing, and therefore, some intrinsic correlation must exist.

On the other hand, existing numerical simulations \cite{lubachevsky1990geometric,torquato2000random,o2003jamming,o2002random,tsekenis2021jamming} and experiments \cite{scott1960packing,bernal1960packing,nowak1998density,pine2005chaos} provide evidence for the existence of RCP, characterized by a consistent packing fraction and other structural features. A commonly used computational method to generate RCP involves the Lubachevsky-Stillinger (LS) algorithm for hard spheres \cite{lubachevsky1990geometric, torquato2000random}, starting from a relatively random initial condition at very low density and then compressing the packing rapidly to achieve higher densities and avoid the crystallization. It is important to note that this process is typically out of equilibrium, and thus the resulting random packings are compression rate dependent. In particular, their final packing fraction decreases as the compression rate. It has been argued that the RCP corresponds to the infinite compression rate. Complementarily, numerical simulations of soft spheres have also observed a Jamming transition at the packing fraction close to $\phi_\text{RCP}$ \cite{liu1998jamming, o2003jamming,o2002random}. Since the methodology of preparing RCP is protocol-dependent, it remains unclear whether there exists a uniquely well-defined RCP \cite{torquato2000random}.

There are several proposals for defining random packing rigorously, each with its own advantages and disadvantages. One approach is to define metrics that evaluate the randomness of packings and then demonstrate that the RCP reaches maximum randomness based on these metrics \cite{torquato2000random,torquato2001multiplicity,atkinson2014existence}. While this approach is straightforward, there is no natural measure of randomness, and different metrics may lead to different types of disorders. Another approach is to define the ensemble of random packings instead of one particular packing\cite{berryman1983random,van1993glass,kamien2007random,song2008phase,mari2009jamming,parisi2010mean,charbonneau2012universal,milz2013connecting,ness2020absorbing,zaccone2022explicit}. This idea is appealing as it is rooted in probability theory and statistical mechanics. Yet, integrating the out-of-equilibrium nature of random packing into an equilibrium statistical mechanics framework presents a significant challenge. The mean-field theory (MFT), developed by the spin glass community, moves towards this goal by introducing intrinsic couplings between the replicas of packings \cite{Mezard1999,parisi2005ideal,parisi2010mean,charbonneau2012universal,mangeat2016quantitative}. However, its validity hinges implicitly on the replica symmetry breaking (RSB) hypothesis \cite{mezard1987spin}. While the RSB has proven effective for infinite-dimensional systems, its validity is currently debated for finite dimensions, even for spin glass models~\cite{fisher1986ordered,nishimori2001statistical,chatterjee2015absence}. Specifically, applying MFT to three-dimensional random packing precludes the possibility of partially ordered structures. Edwards, in contrast, proposed an ensemble of packings sharing the same volume, each with an equal probability of being visited \cite{edwards1989theory,edwards2002granular}. Yet, despite limited efforts\cite{song2008phase,martiniani2017numerical,baule2018edwards}, analyzing the Edwards ensemble remains a challenging task. More recently, there have been proposals to consider the jamming transition as a dynamic phase transition \cite{corte2008random,wilken2021random}, but these approaches also rely on the protocol of generating random packing.

In this paper, we introduce a novel description of random packing, utilizing the ensemble approach outlined above. Our method is inspired by established spin glass models, where the inherent randomness, known as the quenched disorder, drives the system out of equilibrium. For instance, in the Edwards-Anderson (EA) model \cite{edwards1975theory}, the coupling constant between two neighboring spins is randomly drawn from a preset distribution. We suggest a similar strategy can be employed to depict random packing, allowing us to define it explicitly as an ensemble of packing configurations closest to a quenched Poisson random point field. These configurations exhibit maximum randomness in that they resemble uncorrelated random points, i.e., ideal gas, as closely as possible.  Notably, our definition is independent of packing-generating protocols. Moreover, efficient algorithms exist for examining the proposed random packing in two and three dimensions. Our findings indicate a jamming transition at $\phi_J \approx 0.645$, aligning closely with the empirically observed value for RCP. At this point, the three-dimensional random packing configuration transits from local to global rearrangements. Additionally, we identify a novel out-of-equilibrium thermodynamic identity, where the jamming transition corresponds to the emergence of an analogous ``heat". This innovative approach provides a pathway to overcoming the challenges associated with a natural and unambiguous definition of random packing and lays a robust groundwork for future explorations in this intriguing field.

\section*{Problem definition}

Generating RCP typically begins with spheres placed at random positions, a common procedure shared across various protocols. Initially, the packing fraction is kept very small to avoid overlaps between the spheres. As the packing fraction increases, either by compression or increasing particle size, the process implicitly establishes an intrinsic correlation between the initial condition and the final dense packing. The quenched disorder in random packing arises from the random initial condition. This observation motivates us to introduce an ensemble of random packing. For a $d$-dimensional system of $N$ particles with positions $\vec{\r}=\{\r_i\}$, we introduce the partition function
\begin{equation}
Z(\rr)=\int\exp(-
\beta\left(\sum_{i<j} U(\r_i, \r_j)+\epsilon\sum_i (\r_i-\rr_i)^2 \right)d^N\vec\r,
\label{eq:Z0}\end{equation}
where $U(\r_1,\r_2)$ captures the pairwise potential between particles. The initial positions $\vec\rr=\{\rr_i
\}$ are quenched random variables, which we obtain from a Poisson random point field. The coupling constant $\epsilon$ controls the position correlation between the initial condition and final packing. One may consider $\epsilon$ to play a similar role as the compression rate in the LS algorithm. However, the model~(\ref{eq:Z0}) is independent of the particular choice of packing-generating protocols, yet captures the essence of random packing. Choosing $U$ to be the hard-sphere potential, Eq.~(\ref{eq:Z0}) reduces to
\begin{equation}
Z_{HS}(\rr)=\sum_\text{admissible packings}\exp\left(-\beta\epsilon\sum_i (\r_i-\rr_i)^2 \right),
\label{eq:Z}\end{equation}
where ``admissible packings" satisfies the hard-sphere constraint, $(\r_i-\r_j)^2 < \sigma^2$, where $\sigma$ is the diameter of spheres. 
The central physical quantity is the mean square displacement (MSD) $\Delta \equiv \frac{1}{N}\sum_{i=1}^N \overline{\left<(\r_i-\rr_i)^2\right>} = \frac{\partial \overline{F}}{\partial{\beta}}$, where the averaged free energy $\beta \overline {F}=-\overline{\ln Z}$. $\left<\ldots\right>$ and $\overline{\ldots}$ represent ensemble and disorder averages, respectively. To make $\Delta$ dimensionless, we rescale it by $\rho^{2/d}$, where $\rho \equiv N/V$ is the density, and $d$ is the dimensionality of the system. Therefore, we define the dimensionless quantity $\rho^{-2/d} \Delta$ as the MSD, and refer to it as such throughout the article unless otherwise stated. Note, applying the replica trick to Eq.(\ref{eq:Z0}) by taking $m$-copies of replicas and averaging over quenched disorder, leading to a replicated partition function $\overline{Z^m} \sim \int \prod_{a=1}^m d^N \vec\r^a \allowbreak \exp(-
\beta\left(\sum_{a=1}^m \sum_{i<j} U(\r_i^a, \r_j^a) \allowbreak +\frac{\epsilon}{m} \sum_{a<b} \sum_i (\r_i^a-\r_i^b)^2 \right)$. This is the starting point of the MFT approach \cite{parisi2010mean}.  However, our model does not depend on the RSB assumption of the MTF and potentially leads to different physics, as we will discuss below. 

\begin{figure}
  \includegraphics[width=1\linewidth]{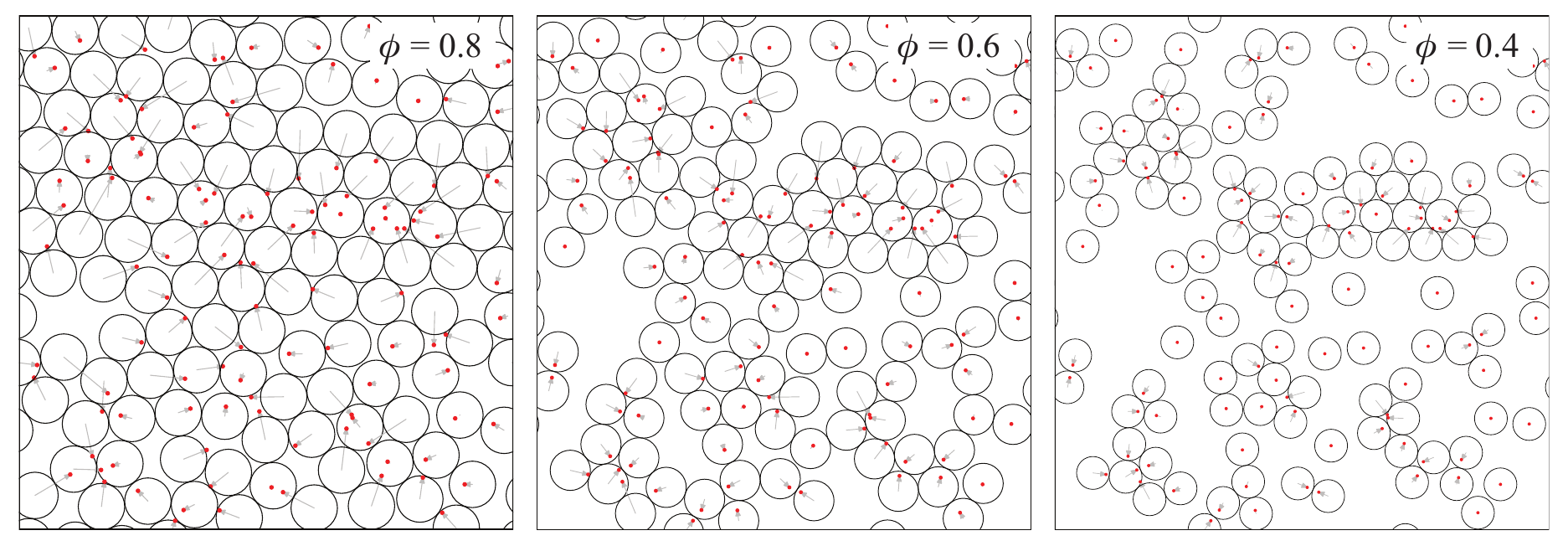}
  \caption{\textbf{Maximally random packing.}  Circle packing with $N=50$ circles, closest to a quenched random Poisson field $\vec\rr$ (red dots) for packing fractions $\phi = 0.8$, $ 0.6$ and $0.4$, respectively. Arrows represent the displacements from the best-matching random points. }
  \label{fig:packing}
\end{figure}

As suggested by existing numerical works, RCP corresponds to close packing for an infinite compression rate, which corresponds to the strong coupling limit $\epsilon \to \infty$. This limit has a simple physical interpretation: the packing configuration is as close as possible to the random initial condition and therefore exhibits maximal randomness. In the following discussion, we will primarily focus on these packings and maintain our consideration within this strong coupling limit, unless stated otherwise. In this limit, Eq.~(\ref{eq:Z}) can be transformed into the following quadratically constrained quadratic program (QCQP) problem,
\begin{equation}
\begin{split}
&\text{miminize } \sum_i(\r_i-\rr_i)^2, \\
&\text{subject to } -(\r_i-\r_j)^2+\sigma^2 \le 0 \text { for all } i < j. 
\end{split}\label{eq:optimization}
\end{equation}
This optimization problem corresponds to finding an admissible packing closest to the quenched random configuration $\rr$, which we refer to as {\it maximally random} packing. Figure~\ref{fig:packing} demonstrates two-dimensional maximally random packings of $N=50$ circles at three different $\phi$ values. The reduced energy density $\frac{1}{N} \lim_{\epsilon\to\infty}E/\epsilon = \Delta$ becomes the MSD, i.e., the objective function of the optimization problem~(\ref{eq:optimization}). Note that optimization problem~(\ref{eq:optimization}) is a well-defined mathematical object, independent of the protocols. Moreover, we define the reduced force on particle $i$ as $\f_i=\lim_{\epsilon\to\infty} \F_i/\epsilon= -\partial_i \Delta = 2(\rr_i-\r_i)$, and the reduced pressure $ \pi = \frac{\phi}{dN} \sum_i \f_i \cdot \r_i$. The reduced force and pressure are also dimensionless quantities, which can be obtained by rescaling the density.

\section*{Jamming transition} 

We develop an efficient algorithm of solving the optimization problem~(\ref{eq:optimization}) (see Methods and Appendix A), and apply it to the three-dimensional packing of $N = 4,000$ particles. We measure the MSD and the reduced pressure as functions of packing fraction $\phi$. With the packing fraction reduced, the random packing undergoes a rearrangement to find a new configuration closer to the quenched random field. If such a rearrangement is local under infinitesimal $\phi$ changes, we can show that (see Appendix C)
\begin{equation}\label{eq:dUdV}
\pi=-(\partial \Delta/\partial(\phi^{-1}))_{\rm{local}},
\end{equation}
which is analogous to the thermodynamic relation $P=-(\partial U/\partial V)_S$. 

\begin{figure}
  \includegraphics[width=1\linewidth]{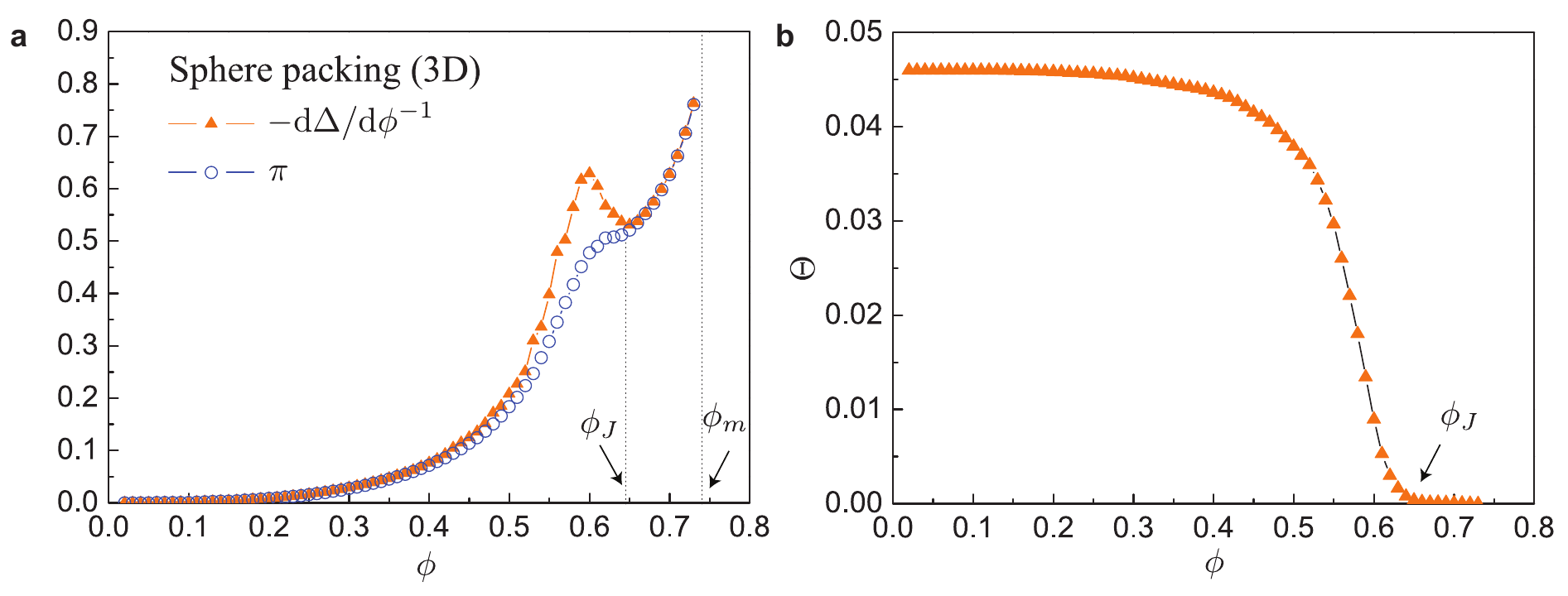}
  \caption{\textbf{Random sphere packing (3D).} (a) The reduced pressure $\pi$ and MSD change $-d\Delta/d\phi^{-1}$ versus the packing fraction $\phi$ for three dimensions for $N=4,000$ spheres, averaged over five realizations of the quenched random points $\vec \rr$. (b) The ``heat'' $\Theta$ as a function of the packing fraction $\phi$. }
  \label{fig:Delta}
\end{figure}

Figure~\ref{fig:Delta} plots the reduced pressure $\pi$ and the MSD change $-\partial \Delta/\partial(\phi^{-1})$ as functions of packing fraction $\phi$ for the three-dimensional packing. We find that Eq.~(\ref{eq:dUdV}) holds only for packing fractions greater than a critical value $\phi_J\approx 0.645$. Below $\phi_J$, Eq.~(\ref{eq:dUdV}) breaks down, implying potential global rearrangements. This finding allows us to introduce an analogous thermodynamic identity
\begin{equation}\label{eq:TI}
\d\Delta = \dbar\Theta - \pi \d\phi^{-1},    
\end{equation}
where the analogous ``heat" $\Theta$ accounts for the discrepancy between the internal energy and work. Integrating Eq.~(\ref{eq:TI}) leads to 
\begin{equation}\label{eq:heat}
\Theta = \Delta - \Delta^{\rm{local}},
\end{equation}
where the work $\Delta^{\rm{local}} \equiv -\int_{\phi_m^{-1}}^{\phi^{-1}} \pi \d\phi^{-1} = \int_{\phi_m^{-1}}^{\phi^{-1}} (\partial \Delta/\partial(\phi^{-1}))_{\rm{local}} d \phi^{-1} $ captures the MSD of local rearrangement. Consequently, the analogous heat corresponds to the excess MSD that accounts for non-local rearrangement. Figure~\ref{fig:Delta}b plots $\Theta$ as a function of $\phi$, showing a phase transition at $\phi = \phi_c$, below which a non-zero heat emerges. 

To understand the emergence of heat better, it is worth noting that the MSD change $-\partial \Delta/\partial(\phi^{-1})$ shows a prominent peak at $\phi\approx 0.59$, resulting in negative slopes between $\phi \approx 0.59$ and $\phi_J$. In contrast, to maintain positive compressibility $\kappa \equiv \phi \d \pi/\d \phi$, the reduced pressure must increase monotonically with $\phi$.  As a result, a non-zero heat is required to resolve this contradiction, implying that global rearrangement is unavoidable. Indeed, this non-zero heat $\Theta$ is related to the instability of local rearrangement when the kissing number $z$ falls below the isostatic bound, i.e., $z < 2d$ (see Appendix D). In simpler terms, the packing remains jammed for packing fraction $\phi >\phi_J$ whereas becomes unjammed for $\phi < \phi_J$. This result indicates a jamming transition at the critical packing fraction $\phi_J$, where the random packing transits from local to global rearrangements. Notably, the critical packing fraction $\phi_J$ aligns closely with the empirically observed $\phi_\text{RCP}$. We thus propose that maximal random packing with the packing fraction $\phi_J$ corresponds to RCP, namely, the ensemble of packings that are closest to the random points while still remaining jammed. 

It is worth noting that the analogous thermodynamic identity~(\ref{eq:TI}) is an out-of-equilibrium phenomenon. Specifically, the heat $\Theta$ does not seem to originate from an entropy change. While there may exist degenerated configurations with a closed distance to the random points $\vec\zeta$, yielding a finite entropy density $s$, the equilibrium relation $\d s = \epsilon \dbar \Theta$ fails at the strong coupling limit $\epsilon\to\infty$. In other words, an infinite entropy density would be required to have a finite heat change, which is unphysical and underscores the out-of-equilibrium nature of random packing.

\begin{figure}
  \includegraphics[width=1\linewidth]{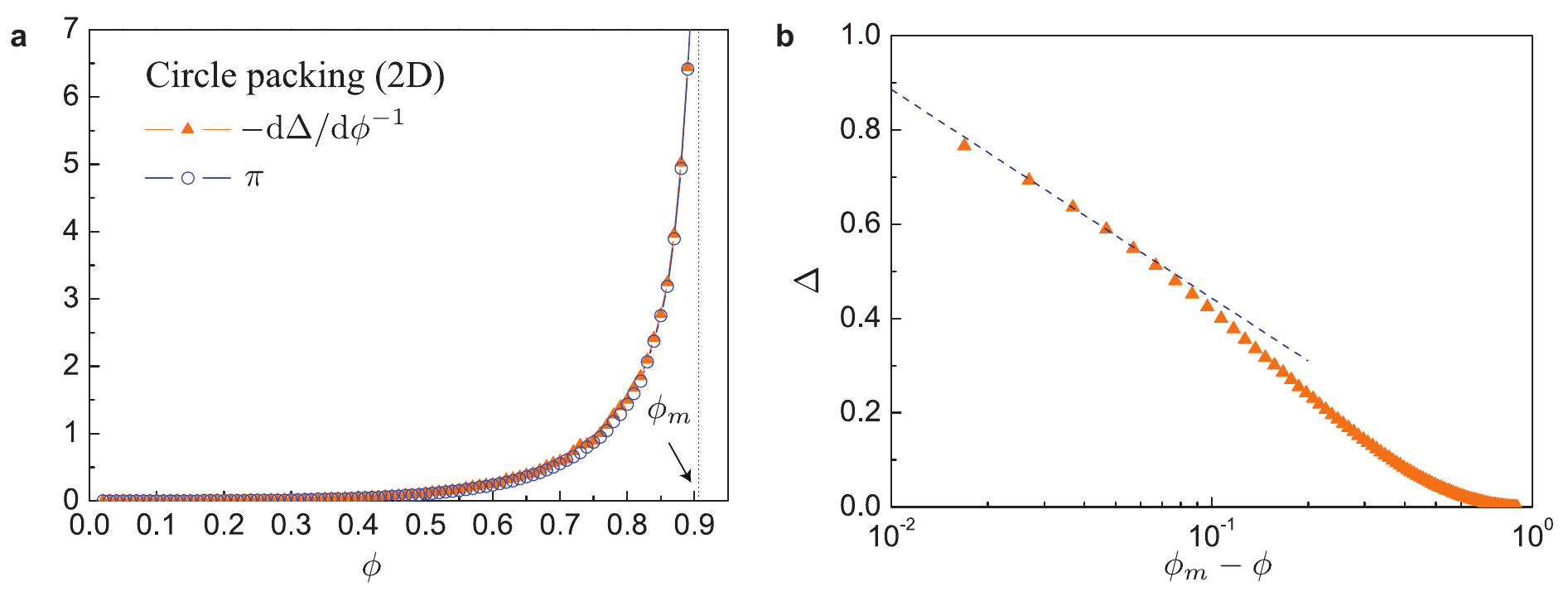}
  \caption{\textbf{Random circle packing (2D).} (a) The reduced pressure $\pi$ and MSD change $-d\Delta/d\phi^{-1}$ versus the packing fraction $\phi$ for $N=4,200$ circles, averaged over five realizations of the quenched field $\vec\rr$. (b) Linear-log plot for $\Delta$ versus $\phi_m-\phi$ for random circle packing, showing a logarithmic divergence (dashed line) when $\phi\to\phi_m$. }
  \label{fig:Delta2d}
\end{figure}

For comparison, we apply the same algorithm to two-dimensional packing. Figure~(\ref{fig:Delta2d})a demonstrates that the reduced pressure closely aligns with the change in MSD across the entire $\phi$ spectrum, in agreement with Eq.~(\ref{eq:dUdV}). This result implies that two-dimensional random packing involves only local arrangement under infinitesimal $\phi$ changes, a characteristic that distinctly contrasts with its three-dimensional counterpart. Moreover, Figure(\ref{fig:Delta2d})b reveals that $\Delta(\phi)$ exhibits a logarithmic divergence when approaching the maximum packing fraction $\phi_m$, i.e., $\Delta(\phi) \sim \log(\phi_m-\phi)$. In contrast, $\Delta(\phi_m)$ is finite in three dimensions, indicating that only a finite distance between the close-packing and random points (see Appendix B). Collectively, these observations highlight a remarkable difference between three-dimensional random packing and its two-dimensional counterpart. However, whether the absence of the jamming transition in two dimensions is associated with the divergence of $\Delta(\phi_m)$ remains an open question. This intriguing matter is set aside for future exploration.

\begin{figure}
  \includegraphics[width=1\linewidth]{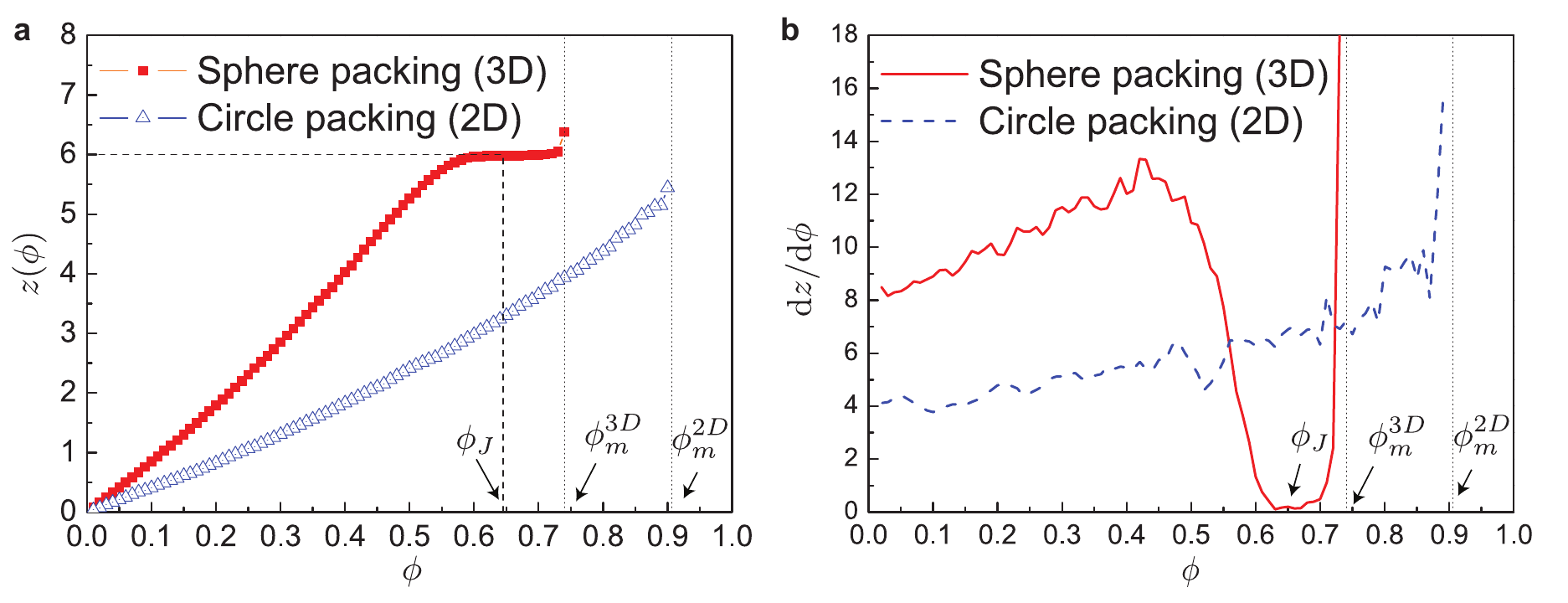}
  \caption{\textbf{Kissing number.} (a) The kissing number $z$ as a function of packing fraction $\phi$ for both random sphere (red, solid squares) and circle (blue, open triangles) packings. (b) The derivative $dz/d\phi$ as a function of packing fraction $\phi$.}
  \label{fig:z}
\end{figure}

The observed jamming transition in the three dimensions is further corroborated by the measurement of the kissing number $z$, which measures the average number of contacts each particle maintains with its neighbors. Figure~\ref{fig:z}a plots $z(\phi)$ as a function of packing fraction $\phi$ for both two and three dimensions. In the three-dimensional setting, we observe a plateau around $z=6$ for random packings with packing fractions near $\phi_J$. In particular, the slope $dz/d\phi$ vanishes when $\phi = \phi_J$ (Fig.~\ref{fig:z}b), thus providing additional evidence of the jamming phase transition at $\phi_J$.  It is plausible that $z-z_c \sim (\phi - \phi_J)^3$ follows a Landau-type cubic relation, but more meticulous numerical simulations are needed to clarify this point in the future. As mentioned earlier, this finding concurs with the Maxwell counting argument \cite{maxwell1870,alexander1998amorphous} that local rearrangement instability arises when $z<2d$ (see Appendix D). In contrast, no such phenomenon is observed in the two dimensions.

Moreover, at a low packing fraction $\phi$, we anticipate the kissing number satisfies $z \sim 2^d \phi$. This prediction is based on an argument that the radial distribution function of an ``ideal" random packing follows $g(r) = \Theta(r-\sigma) + z \frac{\sigma}{d} \delta (r-\sigma)$. Notably, this behavior markedly deviates from simple liquid theory, where $z \neq 0$ solely at the maximum packing fraction $\phi_m$, and from the MFT, where $z \neq 0$ only at the jamming transition point $\phi_J$. Figure~\ref{eq:Z}b validates this hypothesis, showing that $\lim_{\phi \to 0} dz/d \phi = 2^d$.

The analysis presented thus far elucidates the confusion surrounding the definition of random packing, positing that the RCP aligns with the local-global rearrangement transition. A natural query emerges: how does our approach correlate with established theories? As mentioned earlier, applying the replica trick to Eq.~(\ref{eq:Z}) leads to the MFT approach, which is believed to be exact when $d\to\infty$ \cite{parisi2020theory}. Thus, our model should coincide with the MFT predictions at a sufficiently large dimension. However, for the most interesting case of three dimensions, there exists a finite-dimensional effect that cannot be treated by the MFT.

For instance, the MFT approach predicates its theory on the assumption that any weak coupling $\epsilon \to 0^+$ effectively restricts packing to the amorphous phase, thereby circumventing crystallization. This hypothesis may hold in infinite-dimensional scenarios \cite{parisi2020theory}, yet it appears ill-suited for three-dimensional contexts. Our findings suggest that partial crystallization persists even in this strong coupling limit, thereby contradicting the MFT premise. Consequently, the MFT anticipates the MSD, interpretable as the cage size, to vanish at the jamming point. Contrary to this, our numerical simulations demonstrate a finite $\Delta$, underscoring the disparity between the MFT and our approach for three-dimensional random packing.

On the other hand, an intriguing link to the MFT does indeed exist. The MFT rules out the prospect of crystallization. In our terms, it eliminates all local rearrangements that could potentially allow the packing fraction to reach $\phi_m$. As a result, the MFT solely accounts for the excess MSD, obtained by subtracting all local MSD contributions, which is precisely the analogous "heat" defined in Eq.~(\ref{eq:heat}). Indeed, this excess MSD vanishes at the jamming point as shown in Fig.~\ref{fig:Delta}b. Moreover, the peak of $-\d \Delta/\d \phi^{-1}$ occurring at $\phi \approx 0.59$ seems to relate to the so-called dynamic transition predicted by the Mode-Coupling theory (MCT) \cite{van1993glass}. A comprehensive exploration of the connections between our approach and these existing models is left for future studies. Nonetheless, our approach does not rely on assumptions of these existing theories, thus providing an unambiguous picture of three-dimensional random packing, regardless of the validity of MFT or MCT. 

\section*{Amorphous materials}
Although the current work primarily focuses on the strong coupling limit of Eq.~(\ref{eq:Z}), it is worth mentioning that a similar approach can be applied for finite $\epsilon$ values. This is crucial for explaining different observations under various computational and experimental protocols. For instance, in the LS proposal, a range of amorphous close packing with packing fractions from $0.64 - 0.68$ was discovered. In our model, this corresponds to the jamming transitions $\phi_J(\epsilon)$ for different coupling $\epsilon$. The infinitely large compression rate corresponds to the strong coupling limit, with $\phi_J(\epsilon\to\infty) \approx 0.645$ shown in this work. Given that $\epsilon$ is conjugate to $\Delta$, a finite $\epsilon$ will introduce fluctuations to the maximal random packing as $\Delta \to \Delta + O(1/\epsilon)$. A more quantitative study requires the development of a new Monte-Carlo algorithm, a task beyond the scope of this paper.

Our approach can be generalized to any molecular system beyond hard spheres to investigate amorphous materials~\cite{alexander1998amorphous} and structural glass~\cite{bernal1959geometrical,pusey1987observation}. A general protocol for generating amorphous materials typically involves quenching, such as rapid cooling or fast compression. This process generates a coupling between the initial and final states. Generally, one can choose an arbitrary $\vec\rr$ instead of a completely random point field. For instance, it is empirically more relevant to choose $\rr$ as an equilibrium liquid configuration at the temperature when the cooling starts. Although the whole quenching process is a complex dynamic process, our model~(\ref{eq:Z}) offers a simple and natural description that captures the essential off-equilibrium coupling. In general, one would expect the following thermodynamic identity:
\begin{equation}
dU = TdS + \dbar Q_\textrm{oe} - PdV - \epsilon d D,
\end{equation}
where $D = N \Delta$ is the square displacement from the initial state, and $Q_\textrm{oe}$ captures extra off-equilibrium “heat” which potentially relates to a configurational entropy. In the case of hard spheres, we have $dD = Td(S/\epsilon) + \dbar (Q_\textrm{oe}/\epsilon) - (P/\epsilon) dV$, which recovers Eq.~(\ref{eq:TI}) when $\epsilon\to\infty$.  

In this sense, the coupling constant $\epsilon$ as playing a role of a ``temperature" that controls the degree of the non-equilibrium: before quenching, the system is under equilibrium, i.e, $\epsilon = 0$. The quenching process ``heats up" the material by introducing a non-zero coupling, which drives it out of equilibrium. After the quenching stops, the environment sets $\epsilon$ back to zero. However, the material itself is still ``hot" with a non-zero $\epsilon = -\partial U/\partial D$. Then it gradually ``cools down" during contact with the environment, a ``zero-temperature" cold reservoir ($\epsilon=0$). This relaxation process explains qualitatively how aging occurs. This relaxation process potentially provides a physical picture to explain the aging dynamics observed empirically. In some cases, this relaxation takes an infinitely long time, which corresponds to a gapless evolution operator, and thus the system is stuck in an out-of-equilibrium state with a finite $\epsilon$, that is, a true glassy state. Unlike the hard-sphere system, for structural glasses, temperature plays an essential role in competing with the coupling, and a more complicated phase diagram is expected. Overall, our approach provides a novel description of random packing and unveils new insights into amorphous materials, with potential impact on a wide array of related fields.

\begin{acknowledgments}
We are grateful to Yuliang Jin for the insightful discussions and valuable feedback on the draft.
\end{acknowledgments}

\section*{Methods}  
The optimization problem ~(\ref{eq:optimization}) is mathematically well-defined, setting our model apart from prevailing computational approaches. These methods frequently lack rigorous, precise definitions, with the character of random packing often determined by the algorithms themselves. In contrast, any effective algorithm designed to solve Eq.(\ref{eq:optimization}) should produce the same result. However, the non-convex QCQP nature of the problem presents a significant computational challenge in practice, as it is generally recognized as NP-hard \cite{garey1979computers}.

As postulated by MFT, the number of local minima increase rapidly with $\phi$, making the optimization challenging for a high packing fraction. Let us consider the most challenging scenario, where the packing fraction $\phi$ reaches its maximum value, $\phi_m$. With Hassel's validation of the Kepler conjecture~\cite{hales2005proof}, only two potential close-packings remain: the face-centered-cubic (FCC) and hexagonal close-packed (HCP) lattices. As a result, the phase space fragments into $2N!$ disconnected energy barriers, each subject to certain finite-dimensional global symmetries.

Upon initial inspection, the optimization problem~(\ref{eq:optimization}) seems computationally impracticable in this scenario, given that the spheres cannot be smoothly transitioned to alter the packing across $N$ factorial configuration subspaces. Finding the nearest configuration to the random points $\vec\rr$ is then transformed into a combinatorial optimization problem of determining the best matching of random points across the $N$ spheres. This problem is known as the assignment problem or minimum cost perfect matching problem, which involves optimization over all possible $N$ factorial combinations. 

Contrary to intuition, the assignment problem can be efficiently solved within polynomial time \cite{jonker1987shortest}, offering a scalable numerical resolution to ascertain the optimal distance from the random points for maximal close packing. It is worth noting that once the optimal assignment is found, the exchange energy for any pair of particles $i$ and $j$, $(\r_i-\rr_i)^2 + (\r_j - \rr_j)^2 \leq (\r_i-\rr_j)^2 + (\r_j - \rr_i)^2$, or equivalently, $\r_{ij}\cdot \rr_{ij} \leq 0$, where $\r_{ij} \equiv \r_i - \r_j$ and $\rr_{ij} \equiv \rr_i - \rr_j$; otherwise, exchanging labels $i$ and $j$ would lead to a lower energy.

The proposed solution may not be relevant when dealing with highly degenerate, amorphous maximal close packing. In such situations, the challenge extends beyond identifying the best match; it also entails pinpointing the optimal packing configurations among numerous possibilities sharing the same maximal packing fraction. Nevertheless, in scenarios with only a limited number of close-packing configurations, the assignment algorithm can be applied to solve Eq.~(\ref{eq:optimization}) efficiently. This remarkable feature stems from Hales's proof of the Kepler conjecture \cite{hales2005proof}, which helps us determine the ground state of sphere packing. This allows efficient algorithms for computationally investigating random packing, distinguishing it from spin glass models. In contrast, the ground state of the EA model remains a subject of debate \cite{fisher1986ordered,nishimori2001statistical,chatterjee2015absence}.%, with numerical investigations generally limited to small systems.

Given that there are $N!$ energy barriers to search, the optimization process for packing fractions $\phi<\phi_m$ involves two distinct parts: 1) A global optimization across $N!$ barriers using the assignment algorithm to establish the sphere and random point matching, thus determining the globally optimal energy barrier. This step can be initiated with any initial admissible packing, and the assignment algorithm can be applied. 2) A local optimization that rearranges the packing configurations to minimize the MSD within the global optimal energy barrier. We utilize the well-established interior-point method for this purpose \cite{nocedal1999numerical}. We monitor particle permutations to keep $\r_{ij}\cdot \rr_{ij}$ negative to ensure that local optimization remains within the appropriate energy barrier. If this condition is unmet, we swap labels $i$ and $j$ to attain lower energy. 

It is crucial to underscore that the final optimal packing remains independent of the initial interior point. This suggests that any starting choice of admissible packing can be utilized (see Appendix A for details). In contrast, traditional computational approaches often become trapped in a local minimum due to jamming, which prevents the discovery of an admissible packing for packing fraction $\phi > \phi_J$, rendering them dependent on initial conditions and algorithm choice.  However, this does not imply that admissible packings do not exist. Instead, they form a zero-measure set. Diverging from these traditional methods, our proposed algorithm ensures initial admissibility, effectively circumventing the issue of entrapment within local optimization.

\appendix
\section{Numerical Method}\label{sec:algorithm}

In this section, we present an algorithm to address the following optimization problem
\begin{equation}
\begin{split}
&\text{miminize } \sum_i(\r_i-\rr_i)^2, \\
&\text{subject to } -(\r_i-\r_j)^2+\sigma^2 \le 0 \text { for all } i < j. 
\end{split}\label{eq:optimization1}
\end{equation}
for a range of packing fractions $\phi$. We establish our system within a box of size $L$, subject to periodic boundary conditions (PBC). To uphold these conditions, we ensure the center of mass coincides with the box center, i.e., $\sum_i \r_i = 0$. Our proposed algorithm unfolds in several steps:
\begin{enumerate}
\item Begin with $\phi = \phi_m$, where the packing configurations represent maximum close packing -- either FCC or HCP for 3D, and Hexagonal packing for 2D. Generate $N$ quenched points $\rr$ randomly within the box. Construct a complete bipartite graph $K_{N,N}$ between the particles and quenched points, with each edge assigned a weight $w_{ij} = (\r_i-\zeta_j)^2$. Apply the assignment algorithm proposed by Jonker and Volgenant \cite{jonker1987shortest} to determine the optimal match that minimizes the objective function in~(\ref{eq:optimization1}). Because of PBC, a global translation $\rr \to \rr + \r_0$ with a constant $\r_0$ is permissible. This can be addressed by shifting the center of mass of $\rr$ to zero.

\item  Decrease packing fraction: Shift $\phi$ to $\phi - \Delta \phi$ (with a typical choice of $\Delta \phi = 0.1$) by correspondingly reducing the particle diameter $\sigma$. At this new packing fraction, find the new minimum based on the packing optimized at the previous fraction. We start from any admissible packing with the optimal assignment, i.e., the best label matches between the particles and quenched random points such that $\r_{ij}\cdot \rr_{ij} \leq 0$. This can be achieved by the assignment algorithm in Step 1). We then implement the interior-point method to optimize the following auxiliary free energy:
\begin{equation}\label{eq:f}
    f(\R,\sigma) \equiv \sum_i (\r_i-\rr_i)^2-t\sum_{i < j}U(r_{ij}/\sigma),
\end{equation}
where $\R \equiv \{\r_i\}$, $\r_{ij} \equiv \r_i - \r_j$, and $r_{ij} = |\r_{ij}|$. Here, $U(x) = \ln(x^2-1)$, aan auxiliary entropy represented by a logarithmic barrier function that forces the search step to fall inside the hard-sphere constraints, and the parameter $t$ serves as the auxiliary temperature. Readers should not confuse these auxiliary parameters with the physical temperature and free energy defined in Eq.~(1). The interior-point method commences with a relatively high $t$ value, which is iteratively halved until it reaches a sufficiently small value, typically $t = 10^{-10}$. In this sense, the barrier method is similar to simulated annealing, where we start with a sufficiently large auxiliary temperature $t$ and cool down the system by reducing the auxiliary temperature until reaching a minimum. Note that the interior-point method is insensitive to initial conditions, allowing us to use an initial interior point from the previous larger packing fractions, which is admissible with the optimal assignment having been solved, to save computational time. Alternatively, we can also start from any admissible packing, followed by the assignment algorithm. We observe numerically that different choices of initial conditions lead to the same optimal packing.

\item For a given $t$, the optimization~(\ref{eq:f}) is solved using the Limited-memory Broyden–Fletcher–Goldfarb–Shanno (L-BFGS) algorithm \cite{liu1989limited}. To avoid stagnation at a local minimum during the optimization process, we maintain a watch on particle matches to ensure $\r_{ij}\cdot \rr_{ij} \leq 0$. If not, we exchange the labels $i$ and $j$ to achieve a lower energy.

\item Repeat Step 2 iteratively across the entire range of packing fractions.

\end{enumerate}

The forces exerted on the particles, and consequently, the reduced pressure can be evaluated from the optimal configuration. This can be elucidated by observing that, for a fixed parameter $t$, the optimal packing $\R^*$ of Eq.~(\ref{eq:f}) requires the gradient $\g(r,\sigma) \equiv \nabla_{\R} f(\R,\sigma)$ vanish, i.e., $\g(\R^*, \sigma) = 0$, leading to
\begin{equation}
    2(\r_i-\zeta_i) = t \sum_{j} \nabla_{i} U(r_{ij}/\sigma) = \frac{t}{\sigma} \sum_{j} U'(r_{ij}/\sigma) \n_{ij} ,
\end{equation}
where $\n_{ij} \equiv \r_{ij}/r_{ij}$. From this, we can derive the force $\f_i$ acting on particle $i$ as 
\begin{equation}
\f_i = - 2(\r_i-\zeta_i) =\sum_{j} \f_{ij}.
\end{equation}
Here, $\f_{ij}$ is the force of contact between particles $i$ and $j$, as
\begin{equation}
    \f_{ij} =  - f_{ij} \n_{ij},
\end{equation}
with its magnitude
\begin{equation} \label{eq:fij}
    f_{ij} \equiv \frac{t}{\sigma} U'(r_{ij}/\sigma). 
\end{equation}
In the limit as $t$ approaches $0$, only neighboring contacts where $r_{ij} \to \sigma$ remain non-zero. Hence, Eq.~(\ref{eq:fij}) provides a method to compute the pairwise contact force $f_{ij}$ for the optimization algorithm.

\section{MSDs for $\phi = \phi_m$}
\begin{figure}
  \includegraphics[width=1\linewidth]{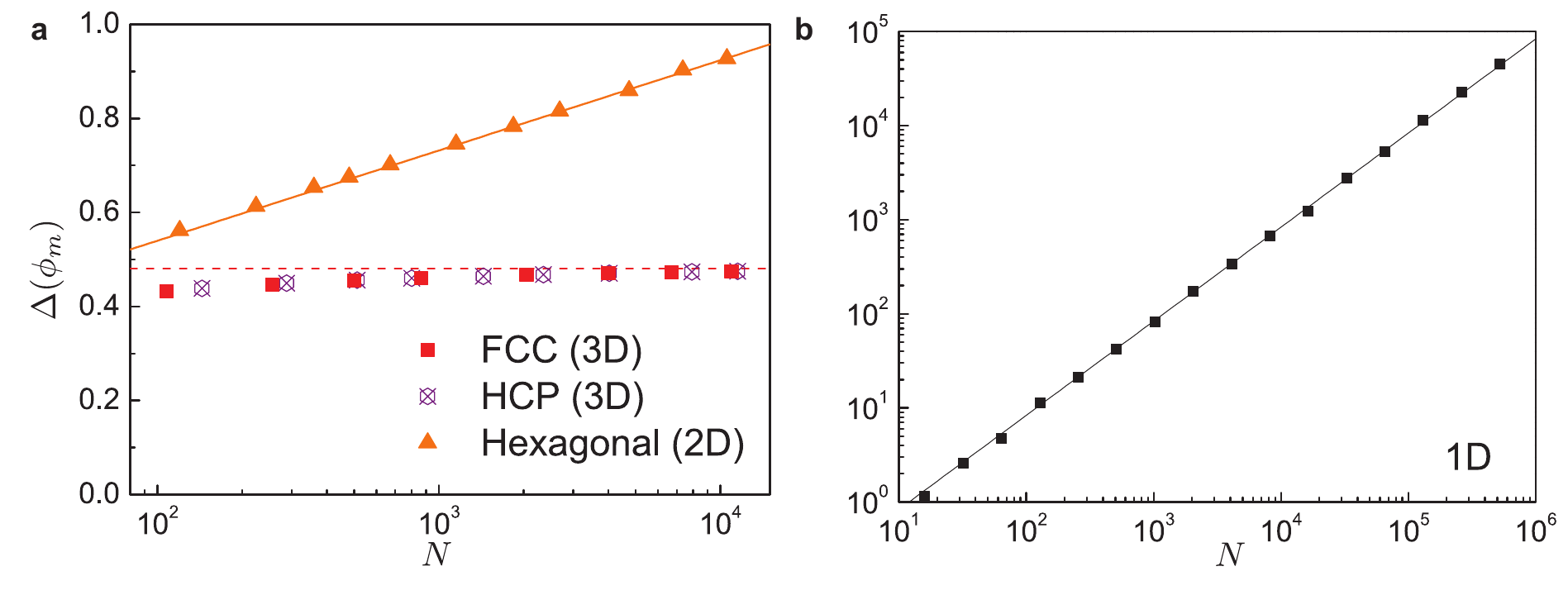}
  \caption{\textbf{Maximum close packing.} The mean square displacement $\Delta(\phi_m)$ as a function of system size $N$, for (a)  FCC (red, solid squares) and HCP (blue, crossed squares) sphere packings,  and two-dimensional hexagonal packing (orange, solid triangles), and for (b) one-dimensional packing (black, solid squares). The orange and black lines represent $\phi_m \sim \ln N$ for 2D, and $\phi_m \sim N/12$ for 1D, respectively. }
  \label{fig:scaling}
\end{figure}

Figure~\ref{fig:scaling} investigate the finite scaling of MSD as a function of the system size $N$. Our results show that, for $d < 3$, $\Delta(\phi_m)$ diverges with increasing $N$, indicating a delocalized phase. In particular, for $d = 1$, the MSD scales linearly with size, $\Delta(\phi_m) \sim N/12$ for the periodic boundary condition and $\sim N/6$ for the open boundary condition. For two dimensions, our numerical results suggest a logarithmic divergence, i.e., $\Delta(\phi_m) \sim \ln N $. In contrast, $\Delta(\phi_m)$ is finite in three dimensions for both HCP and FCC packings. This result is rather surprising because one would expect a large configurational change from the most ordered packings (FCC/HCP) to the most disordered ordered ones (Poisson). However, we find the opposite: a bounded displacement is sufficient to transform one to the other, implying that all three-dimensional packings are close to each other. It appears that $\Delta \sim (\phi_m-\phi)^{2-d} + O(1)$, suggesting that $d = 2$ is the lower critical dimension of the delocalization, which appears linked to the jamming transition.

%There are also many other interesting directions for future studies. For instance, the MSD near the maximum packing fraction $\phi_m$ is also worthy of study in its own right. As we have demonstrated, 

\section{Local rearrangement}\label{sec:local}
In this section, we  derive the relation between the reduced pressure $\pi$ and the change of MSD $\d \Delta/\d \phi^{-1}$ under a local rearrangement. By local rearrangement, we mean that under a small change of diameter $\sigma$ and thereby the packing fraction $\phi$, the optimal packing configuration only deforms slightly.

Starting with a fixed diameter $\sigma$ and the corresponding optimal packing $\R^*$, we expand $f(\R,\sigma)$ around $\R^*$,
\begin{equation}
 f(\R,\sigma)=f(\sigma)+\g(\R^*,\sigma)\cdot\delta\R^*+\frac{1}{2}(\delta\R^*)^T \H(\R^*,\sigma)\delta\R^*+O((\delta\R^*)^3),
\end{equation}
where $f(\sigma) \equiv f(\R^*,\sigma)$, the gradient 
$\g(\R,\sigma) \equiv \nabla_{\R} f(\R,\sigma)$ satisfies $\g(\R^*, \sigma) = 0$, and the Hessian $\H(\R,\sigma) \equiv  \nabla_{\R} \otimes  \nabla_{\R} f(\R,\sigma)$.

Considering a small perturbation $\sigma \to \sigma + \delta \sigma$, and $\R^* \to \R^* + \delta\R^*$, where $\delta\R^*$ only undergoes a local rearrangement and is of the same order as $\delta \sigma$, the new minimum is $f(\sigma + \delta \sigma)  \equiv f(\R^* + \delta\R^*, \sigma + \delta \sigma)$. Expanding $f$ around $(\R^*, \sigma)$, we have
\begin{equation}\label{eq:fsig}
f(\sigma+\delta \sigma )=f(\sigma)+\partial_\sigma f(\R^*,\sigma)\delta \sigma+\partial_\sigma \g(\R^*,\sigma)\cdot\delta\R^*\delta \sigma +\frac{1}{2}\left(\partial_\sigma^2 f(\R^*,\sigma)\delta \sigma^2+(\delta\R^*)^T \H(\R^*,\sigma)\delta\R^* \right)+O(\delta\sigma^3)
\end{equation}
Minimizing $ f(\R^* +\delta\R^*,\sigma+\delta \sigma ) $ requires $\partial_{\delta\R^*} f(\R^* +\delta\R^*,\sigma+\delta \sigma ) = 0$, leading to 
\begin{equation} \label{eq:dR}
\delta\R^*=-\H^{-1}(\R^*,\sigma)\partial_\sigma \g(\R^*,\sigma) \delta \sigma
\end{equation}
Substituting into Eq.~(\ref{eq:fsig}), we obtain
\begin{equation}\label{eq:fTalor}
    f(\sigma + \delta \sigma) =f(\sigma)+ \frac{\d f(\sigma)}{\d \sigma} \delta \sigma +\frac{1}{2}\frac{\d^2 f(\sigma)}{\d\sigma^2}\delta \sigma^2+O(\delta \sigma^3),
\end{equation}
where
\begin{subequations}
\begin{equation}\label{eq:f1}
\frac{\d f(\sigma)}{\d \sigma}=\partial_\sigma f(\R^*, \sigma),
\end{equation}
\begin{equation}\label{eq:f2}
\frac{\d^2 f(\sigma)}{\d\sigma^2}=\partial_\sigma^2 f(\sigma,\R^*)-\partial_\sigma \g(\R^*,\sigma)^Th(\R^*,\sigma)^{-1}\partial_\sigma \g(\R^*,\sigma).
\end{equation}
\end{subequations}

Note that from Eq.~(\ref{eq:f}), we have
\begin{equation}\label{eq:RF}
\R\cdot\nabla_{\R} f(\R,\sigma) = \R \cdot \F - \sigma \partial_\sigma f(\R,\sigma),
\end{equation}
where $\F \equiv \{\f_i\}$. The validity of this identity is independent of the choices of $t$ and $U$ functions, and thus holds also in the optimization limit $t\to 0$. Setting $\R = \R^*$ and $\F = \F^*$ and taking the limit $t\to 0$, we have $\lim_{t\to 0} f(\R^*,\sigma) = N \Delta(\sigma)$. Substituting Eq.~(\ref{eq:f1}) into Eq.~(\ref{eq:RF}) and noting that the gradient $\nabla_{\R} f(\R,\sigma)$ vanishes at $\R^*$, we obtain
\begin{equation}\label{eq:RF2}
   \frac{1}{N} \R^* \cdot \F^* = \sigma \frac{\d \Delta}{\d\sigma} = \phi d \frac{\d \Delta}{\d \phi}.
\end{equation}
Recall the definition of the reduced pressure
\begin{equation}
    \pi \equiv \frac{\phi}{d N} \sum_{i<j} \r_{ij}\cdot \f_{ij} = \frac{\phi}{d N} \R^* \cdot \F^*,
\end{equation}
where the second equality arises from the Virial theorem, $
    \R^*\cdot\F^* = \sum_i \r_i \cdot \f_i = \frac{1}{2} \sum_{ij} \left( \r_i \cdot \f_{ij} + \r_j \cdot \f_{ji}\right) = \sum_{i<j} \r_{ij} \cdot \f_{ij}
$. Substituting into Eq.~(\ref{eq:RF2}), we obtain the relation between $\pi$ and $\Delta$ for the local rearrangement
\begin{equation}
   \pi = -\frac{\d \Delta}{\d \phi^{-1}}.
\end{equation}
This completes our proof.

\section{Jamming transition}\label{sec:jamming}
In this section, we show the condition under which local rearrangement is valid and its association with the jamming transition. The validity of Taylor expansion in Eq.~(\ref{eq:fTalor}) requires a well-defined second derivative
\begin{equation}\label{eq:f2}
\sigma^2\frac{\d^2 f(\sigma)}{\d\sigma^2}=\sigma^2\partial_\sigma^2 f(\sigma,\R^*)-\G^T \H^{-1}\G.
\end{equation}
where $\G \equiv \sigma \partial_\sigma \g(\R^*,\sigma)$. By respectively applying $\nabla_{\R}$ and $\partial_\sigma$ to both sides of Eq.~(\ref{eq:RF}), we obtain 
\begin{subequations}
\begin{equation}
   \H \R^*  = (\F^* - 2 \R^*)-\G,
\end{equation}
\begin{equation}
     \R^*\cdot \G  = - \R^*\cdot\F^* - \sigma^2\partial_\sigma^2 f(\sigma,\R^*).
\end{equation}
\end{subequations}

Upon integrating these results, we find
\begin{equation}\label{eq:f2}
 \sigma^2\frac{\d^2 f(\sigma)}{\d\sigma^2} =  (2 \R^* -\F^* )\cdot  \q  - \R^*\cdot\F^* ,
\end{equation}
where $\q \equiv \H^{-1} \G$. One can write Eq.~(\ref{eq:f2}) as 
\begin{equation}\label{eq:f3}
 \sigma^2\frac{\d^2 f(\sigma)}{\d\sigma^2} =  \sum_{i<j} (2N^{-1} \r_{ij}- \f_{ij})\cdot \q_{ij} - \f_{ij}\cdot \r_{ij},
\end{equation}
where $\q_{ij} \equiv \q_i - \q_j$.

Upon taking the limit as $t\to 0$, Eq.~(\ref{eq:fij}) implies that $U' \sim t^{-1}$ as $r_{ij} \to \sigma$, ensuring that $\F^*$ is finite. Conversely, both $\G$ and $\H$ diverge because they involve $U''$, which exhibits a stronger divergence as $r_{ij} \to \sigma$. To have a finite $f''(\sigma)$, we require
\begin{equation}\label{eq:q}
  \q^* \equiv \lim_{t\to 0} \q = \lim_{t\to 0} \H^{-1} \G ,
\end{equation}
to be well-defined. Indeed, Equation~(\ref{eq:dR}) suggests
\begin{equation}
\frac{\delta\R^*}{\delta \sigma} =-\sigma^{-1} \q^*. 
\end{equation}
Therefore, the local rearrangement requires $\q$ to be finite. To illustrate this more clearly, we find
\begin{equation}
    \G_i = - \f_i - \frac{t}{\sigma^2} \sum_j U''(x_{ij}) \r_{ij}
\end{equation}
where $x_{ij} \equiv r_{ij}/\sigma$. Similarly, we find
\begin{equation}
    \H_{ij} = \frac{t}{\sigma^2} \left( -U''(x_{ij}) \n_{ij} \otimes \n_{ij} + x_{ij}^{-1} U'(x_{ij}) (\n_{ij} \otimes \n_{ij}-I_d) \right) 
\end{equation}
is applicable for off-diagonal elements $i \neq j$, where $I_d$ is the $d$-dimensional identity matrix, and
\begin{equation}
    \H_{ii} = 2 I_d - \sum_{j \neq i} \H_{ij}
\end{equation}
is used for diagonal elements. We assume that $U''$ diverges as $(t\epsilon(t))^{-1}$ for small $t$, where the function $\epsilon(t)$ captures the asymptotic behavior such that $-\lim_{t\to 0}\frac{\epsilon(t)t}{\sigma^2} U''(x_{ij}) = u_{ij}$ is finite for neighboring contacts. For instance, for the logarithmic barrier function $U(x) = \ln(x-1)$, $U'(x) = \frac{1}{x-1}$ and $-U''(x) = \frac{1}{(x-1)^2}$, Eq.~(\ref{eq:fij}) implies that only the neighboring contact $r_{ij} = \sigma + \frac{1}{f_{ij}}t + O(t^2)$ contributes. Additionally, setting $\epsilon(t) = t$, we find that $- \lim_{t\to 0} \epsilon(t)\frac{t}{\sigma^2} U''(x_{ij}) = f_{ij}^2 = u_{ij}$. The argument below will be maintained for a general $U$ form, as the specific choice of the barrier function does not alter the result. Equation~(\ref{eq:q}) is consequently transformed into
\begin{equation}\label{eq:Hq}
  \H^* \q^* = \G^*,
\end{equation}
where $\H^*_{ij} \equiv \sigma^{-2} u_{ij} \r_{ij} \otimes \r_{ij} $ with $\H^*_{ii} \equiv -\sum_{j\neq i} \H^*_{ij}$, and $
\G^*_{i} \equiv \sum_{j} u_{ij} \r_{ij}$. We expand   
\begin{equation}
    \H^* = \sum_{ij} h_{ij} |\v^{(ij)}\rangle\langle \v^{(ij)}|,
\end{equation}
in terms of the basis $\{\v^{(ij)}\}$, where the coefficient $h_{ij} \equiv  -\sigma^{-2} u_{ij} $, and $\v^{(ij)} = ( \e^{(i)}-\e^{(j)} ) \otimes (\r_i-\r_j) $ are $dN$-dimensional vectors. Here, $\e^{(i)}$ is the $N$-dimensional vector with a unit at position $i$ and zeros elsewhere. The number of bases $\{\v^{(ij)}\}$ is equivalent to the number of contacts $zN/2$, with $z$ representing the kissing number.

The basis set $\{\v^{(ij)}\}$ is not orthogonal and possesses a $d$-dimensional non-trivial kernel $\T = \sum_i \e^{(i)} \otimes \r_0$ with an arbitrary $d$-dimensional vector $\r_0$, such that $\T \cdot \v^{(ij)} = 0$. This property corresponds to the global translational invariance of the PBC, i.e., $\r_i \to \r_i + \r_0$. Consequently, $\H^*$ has a maximum rank $d(N-1)$ and is not directly invertible. However, Eq.~(\ref{eq:Hq}) does offer a solution since $\G^*$ resides outside of the null space, i.e., $\G^* \cdot \T = 0$. In fact, it encompasses a $d$-dimensional solution space, as if $\q^*$ serves as a solution, then $\q^* + \T$ is also a valid solution. This ambiguity, however, does not affect Eq.~(\ref{eq:f3}) as $\q_{ij}$ is invariant under the global translation.

Eq.~(\ref{eq:Hq}) allows solutions when the rank of $\H^*$ reaches its maximum value $d(N-1)$. Given that there are $zN/2$ bases, it requires $ zN \geq 2d(N-1)$. In the thermodynamic limit of $N\to \infty$, we recover the Maxwell counting argument
\begin{equation}
   z \geq 2d,
\end{equation}
which provides a necessary condition for the validity of the local rearrangement. To ensure its sufficiency, we need the rank of these $zN/2$ base vectors equal to $d(N-1)$, a condition that depends on the configuration $\R^*$. Our numerical results, however, indicate that this condition is likely met, at least statistically. Thus, the isostatic condition corresponds to the critical value $z_c = 2d$ at the jamming transition. Any value below this threshold triggers a global rearrangement of the system, leading to its unjammed state.

% \begin{acknowledgments}
% C.S. was supported partially by the National Science Foundation under Grant IBSS-1620294, the Institute of Education Sciences under Grant R324A180203, and the National Institutes of Health under Grant R01DC018542. 

% J.L. and S.H. contributed equally to the work.  
% \end{acknowledgments}

\bibliography{ref}% Produces the bibliography via BibTeX.

\end{document}